# Frequency Dependent Dynamical Electromechanical Response of Mixed Ionic-Electronic Conductors

A.*N. Morozovska,*[1,*] *E.A. Eliseev,*[1,2] *S.L. Bravina*[3]**,** *Francesco Ciucci*[4†]**,** *G.S. Svechnikov*[1]*, Long-Qing Chen*[5] *and S.V. Kalinin*[6]

[1] Institute of Semiconductor Physics, National Academy of Science of Ukraine, 41, pr. Nauki, 03028 Kiev, Ukraine

[2] Institute for Problems of Materials Science, National Academy of Science of Ukraine, 3, Krjijanovskogo, 03142 Kiev, Ukraine

[3] Institute of Physics, National Academy of Science of Ukraine, 46, pr. Nauki, 03028 Kiev, Ukraine

[4] The Hong Kong university of Science and Technology. Department of Mechanical Engineering. Department of Chemical and Biomolecular Engineering. Clear Water Bay, Kowloon, Hong Kong

[5]Department of Materials Science and Engineering, Pennsylvania State University, University Park, Pennsylvania 16802, USA

[6] The Center for Nanophase Materials Sciences, Oak Ridge National Laboratory, Oak Ridge, TN 37922

---

[*]morozo@i.com.ua

[†]email mefrank@ust.hk

**Abstract**

Frequency dependent dynamic electromechanical response of the mixed ionic-electronic conductor film to a periodic electric bias is analyzed for different electronic and ionic boundary conditions. Dynamic effects of mobile ions concentration (stoichiometry contribution), charge state of acceptors (donors), electron concentration (electron-phonon coupling via the deformation potential) and flexoelectric effect contribution are discussed. A variety of possible nonlinear dynamic electromechanical response of MIEC films including quasi-elliptic curves, asymmetric hysteresis-like loops with pronounced memory window and butterfly-like curves are calculated. The electromechanical response of ionic semiconductor is predicted to be a powerful descriptor of local valence states, band structure and electron-phonon correlations that can be readily measured in the nanoscale volumes and in the presence of strong electronic conductivity.

**Keywords:** thin films of ionic semiconductors, dynamic electromechanical response, deformation potential, flexoelectric effect.

# 1. Introduction

Materials with dual electronic and ionic conductivity, referred to as mixed electronic ionic conductors (MIECs) are broadly used in energy related applications such as batteries [1, 2], sensors [3, 4] and fuel cells [5, 6], as well as electronic device applications including memristive, and electroresistive memory and logic devices [7, 8]. Beyond these applications, ionic and electrochemical effects can heavily contribute to the operation of ferroelectric devices [9, 10] and capacitors, including ferroelectric fatigue [9,10], ferroelectric resistive switching [11], ferroelectric gate devices [12], or spurious observations of ferroelectricity in centrosymmetric materials in bulk [13] or SPM geometries [14, 15, 16], piezoresistive phenomena [17], and exotic memory and transport effects in nano- and molecular electronic devices [18]. Recently, ionic phenomena are considered as an origin of unique properties of $LaAlO_3$-$SrTiO_3$ interfaces [19]. Many oxides such as manganites, cobaltite, and ferrites, are both extensively studied in condensed matter physics community [20, 21] and are used in energy applications, pointing at the possible role of ionic phenomena in classical physical studies. The multitude of ionic phenomena in nanoscale systems necessitates the development of comprehensive measurement strategies applicable for nanoscale materials in the form of capacitor-like device structure and scanning probe microscopy (SPM).

Understanding of physical and electrochemical phenomena in these materials necessitates development of measurement techniques addressing local valence states and electrochemical functionality and their response to external bias and chemical stimuli on the local scale. Significant progress in this direction has been achieved with the advent of electron-microscopy based electron energy loss spectroscopy (EELS) imaging [22] and synchrotron based X-ray measurements. However, the understanding of these systems can be considerably extended if these studies can be extended to local probing of functionality on a single grain, defect, or domain wall level, combining the broad spectrum of capabilities of conventional electrochemical characterization techniques and high spatial resolution of electron and scanning probe microscopies.

The applicability of traditional electrochemical measurements based on the Faradaic current detection is necessarily restricted to the 1 -50 micron length scale due to the electronic current detection limits [23, 24]. The comprehensive analysis [25] of recent efforts in extending the electrochemical charge-discharge [26, 27, 28, 29] or impedance spectroscopy [30, 31, 32] methods to SPM environment suggest that these studies are possible only when the process is catalyzed at and around the tip surface junction [33]. At the same time, when tip or surface material are active part of the ionic or electrochemical process, corresponding current cannot be probed directly and the progression of reaction can be ascertained only through static changes in surface topography, Raman signature, etc (e.g. for tip-induced nanooxidation [34, 35], or deposition of carbon [36], semiconductors [37] or metals [38, 39]). This limitation stems both from the smallness of Faradaic currents and presence of surface leakage currents (dc detection) and stray capacitances of the measurement circuits (ac detection).

An alternative approach for probing ionic and electrochemical processes in solids is based on electromechanical and chemo-mechanical coupling. The latter correspond to chemical expansivity measurements of volume vs. chemical potential of volatile component using macroscopic dilatometric test systems or scattering methods, and are now broadly used in solid state ionic community. In electromechanical methods, potential induced by mechanical stimuli applied to MIEC (direct effect) or mechanical response induced by electrical stimuli (inverse effect) is detected. In particular, this approach offers the advantage of direct implementation in capacitor-like structures with interferometric or vibrometric detection of associated surface displacements, or implementation in the SPM set up. An example of this approach is Electrochemical Strain Microscopy (ESM) [40, 41] in which the periodically biased conductive SPM tip concentrates electric field in a small volume of the material, resulting in redistribution of mobile ions through diffusion and electromigration mechanisms. The associated changes in molar volume and strains results into periodic surface displacement detected by an SPM tip, somewhat similar to the Piezoresponse Force Microscopy (PFM) of piezoelectric and ferroelectric materials [42, 43, 44, 45, 46].

In systems with large background conductivity, the bias-induced electromechanical process can be separated into the interfacial reaction process and subsequent diffusion of the chemical species through the lattice. This problem then reduces to the solution of linear diffusion or coupled diffusion-strain equations, and is well developed in the context of electrochemical storage and metallurgical systems [47, 48, 49, 50, 51, 52]. However, in MIEC systems, the response will be determined by coupled ionic and electronic motion, giving rise to significantly more complex coupled diffusion-migration problem.

To the best of our knowledge the **dynamical local electromechanical properties** of MIECs was poorly studied theoretically [23, 24, 40, 53, 54], in contrast to the theory of their dynamical current-voltage response, that is well elaborated (see e.g. classical papers of Riess et al. [7, 8] and Strukov et al. [55] and refs therein). The basis of theory of **static local electromechanical properties** of MIECs is presented in [56].

The paper is organized as following. The electromechanical and transport phenomena in MIECs is summarized in Section 2. Section 3 contains basic equations and boundary conditions for dynamic local electromechanical response calculations. Results of the dynamic response calculations are presented and analyzed in the Section 4. The linear response is analyzed in the Subsection 4.1. Nonlinearity effect on the response is analyzed in the Subsections 4.2-3. These sections are followed by the brief discussion and summary remarks in Section 5.

## 2. Electromechanical and transport phenomena in MIECs
### 2a. Electromechanical coupling in MIECs

Electromechanical coupling in MIECs is controlled by relationships between molar volume and local field, carrier concentration, and ionic concentration. Well-known effect of the stoichiometry on the local strain is the (often linear) dependence of lattice constants on the composition of solid solution (Vegard law of chemical expansion). The deviations from Vegard law are typically indicative of non-trivial physical phenomena including phase separation, metal insulator transitions, and thus are intrinsically linked to the fundamental physics of the material. Recent experimental studies of correlated oxides including ceria, titanates, ferrites, cobaltites, nikelates analyze chemical expansion effects as related to the oxygen vacancies appearance and migration. Adler et al [57, 58, 59] analyzed the temperature and oxidation-state dependence of lattice volume in $La_{1-x}Sr_xCo_yFe_{1-y}O_{3-\delta}$ ceramics in terms of thermal and chemical expansion. Similar effect of lattice expansion due to the oxygen non-stoichiometry was observed earlier by the different authors (see e.g. Refs. [60, 61, 62]). Bishop et al [63] studied the chemical expansion and oxygen non-stoichiometry of undoped and Gd-doped cerium oxide exposed to different partial pressures of oxygen and found that the contribution to a chemical expansion could be attributed to the larger crystal radius of cerium $Ce^{3+}$ compared to the cerium $Ce^{4+}$. Phenomenological models accounting for the difference in the dopant cation radius and charge as well as the formation of oxygen vacancies have been used to explain experimental results for fluorite-structure oxides [64, 65] assuming linear relations resembling Vegard law. Lankhorst et al. [66] established the relationship between defect chemistry, oxygen nonstoichiometry, and electronic properties (i.e. Fermi level position) in MIEC $La_{1-x}Sr_xCoO_{3-\delta}$.

The second effect leading to electromechanical coupling in oxides are the *electron-phonon coupling* via deformation potential [56, 67]. Strong *electron-phonon coupling* associated with the *local Jahn-Teller distortion* was proposed as a possible origin of this very unusual behaviour of materials with transition-metal ions [20]. Coupling between orbital occupancy and the Jahn-Teller distortion can play a major role as a driving force of symmetry breaking, because the orbital occupation may strongly couple to the lattice (anion distortion) in some cases [20]. Jahn-Teller distortions are typical for correlated oxides with partially filled d-orbitals (e.g. $t_{2g}$ and $e_g$ for octahedral and tetrahedral coordinates) such as $La_{1-x}Sr_xMnO_{3-\delta}$ [68], $La_{1-x}Sr_xCoO_{3-\delta}$ and even $SrFe_xTi_{1-x}O_{3-\delta}$ [69]. The band gap of $La_{1-x}Sr_xMnO_{3-\delta}$ (~ 1 eV) is mainly determined by the collective Jahn-Teller distortion [70]. Since the deformation potential is directly related with the band gap in the narrow gap semiconductors, Fermi level in (half) metals as well as with the charge gap in correlated metal-insulators [71, 72], the electromechanical response of correlated oxides like p-$La_{1-x}Sr_xMnO_{3-\delta}$ could provide the important information about the local band structure and Jahn-Teller distortions.

Using paraelectric $SrTiO_3$ film as a model material with well known electromechanical, electronic and electrochemical properties, we have previously evaluated the contributions of electrostriction, Maxwell stress, flexoelectric effect, deformation potential and compositional Vegard strains caused by mobile vacancies (or ions) and electrons to the *static* electromechanical response [67]. Furthermore, in Ref. [56] we developed a **thermodynamic** approach that allows evolving theoretical description of linear mechanical phenomena induced by the electric fields (electro-mechanical response) in solid state ionics towards analytical theory and phase-field modeling of the MIECs in different geometries and under varying electrical, chemical, and mechanical boundary conditions. These results motivate to continue our theoretical study on **dynamic effects** in the present manuscript.

*2b. Transport in MIECs*

To explore the ***dynamic electromechanical response***, the knowledge of the changes in electrochemical potentials of electrons and ions induced by bias are required. This coupling was extensively explored in the context of transport modelling in MIECs is the *dc* and *ac* regimes in the framework of the Boltzmann-Planck-Nernst (**BPN**) approximation for chemical potential and/or Debye linear screening theory assuming constant conductivity. Below, we provide a ***brief overview of recent theoretical studies*** of coupled diffusion-migration transport in MIECs and factors contribution to the mechanical effects in MIECs.

Gil et al [7, 8] analyzed current-voltage characteristics of metal/semiconductor film/metal structures assuming small variations of holes (electrons) and mobile acceptors (donors) concentrations, valid the analytical solution were derived in linear BPN approximation. Svoboda and Fischer [73] considered the internal stress relaxation in thin films due to the vacancies diffusion only, Tangera et al analyzed [74] the distribution of one type space charge in oxide film between blocking electrodes, but the current was regarded absent. Using boundary conditions involving the discharge rate for conductance currents at the interfaces, as proposed by Chang and Jaffe [75], Macdonald [76] considered mobile electrons and holes, while supposing the charged ions uniformly distributed independently on applied voltage, supposed small in comparison with thermal energy. Chen [77] compared two approximate models (local electro neutrality and constant electric field) with numerical solution of BPN equations for fluxes of electrons and oxygen vacancies. Jamnik and Maier [78] proposed equivalent circuit for the model system with constant ionic conductivity. Franceschetti and Macdonald [79] considered exact solution of the BPN equation for steady state of the system with holes, electrons and immobile charged defects. Also they numerically simulated transient currents as system response to step changes of applied bias.

Recently Riess and Maier [80] proposed an extension of linear irreversible thermodynamics to the case of large driving forces expressing the current via nonlinear function of the drop of electrochemical potential over the "local" hopping distance.

Ciucci et al [81] developed a numerical and analytical framework for the study of small bias response and electrochemical impedance of MIECs. These authors linearized the Poisson Nernst-Planck equations and analyzed 10 μm samples of heavily doped MIEC. Therefore those results (impedance equations) were derived in the assumptions of local electroneutrality. The use of the neutral approximation is questionable when the MIECs' lengthscale falls in the nanometer range.

**3. Basic equations and boundary conditions for dynamic electromechanical response**

Here we consider planar configuration, corresponding to the top electrode on the MIEC film. These structures are now actively fabricated for impedance based studies [82, 83], and can also be used for focused X-ray (e.g. Ref. [84] for ferroelectric materials and Ref. [85] for semiconductor nanostructures) interferometric and vibrometric detection. We further note that fully 1D case implies no lateral current and ionic transport from the edges (or completely blocking lateral walls), whereas deposited electrode can allow for lateral transport at the edges. This affects conservation laws for electrons and ionic species. The SPM experiment with localized tip corresponds to the limiting case of very small electrode.

Geometry of the considered asymmetric heterostructure electrode / possible gap / ionic semiconductor film / substrate electrode" is shown in **Fig. 1a.**

**Figures**

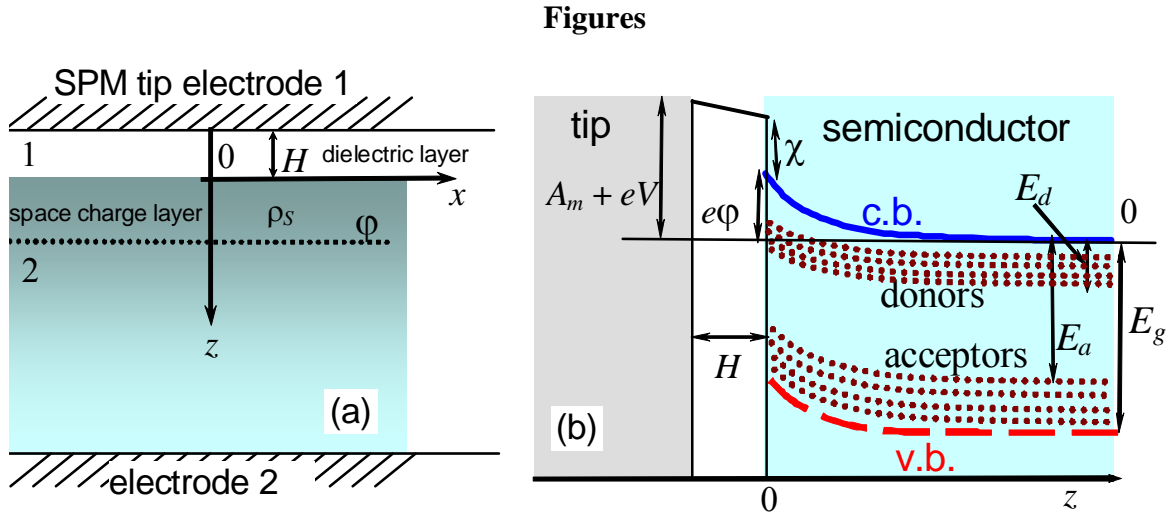

**Fig. 1.** (Color online). (a) Geometry of the considered asymmetric heterostructure "tip electrode/ gap / ionic semiconductor film / substrate electrode". (b) Schematic bend structure at z=0: $U = A_m/e + V$, where $A_m$ is the work function difference, $V(t) = V_0 \sin(\omega t)$ is the voltage difference applied to the tip electrode at $z = -H$, $\varphi$ is the electric potential, $\chi$ is the electron affinity in semiconductor.

Electric potential *V* is applied to the top planar microelectrode. The substrate electrode located at *z=h* is grounded, *V* = 0. The semiconductor film of thickness *h* is regarded thick enough to have a continuous band structure [**Fig. 1b**]. The existence of vacuum or air dielectric gap of thickness *H* between the charged top microelectrode and the MIEC film is also possible, mirroring models for imperfect contact or dead layer in ferroelectric materials [86, 87].

Here, free electrons in the conductive band (*n*) and holes (*p*) in the valence band are considered, which quasi-levels can be different and coordinate dependent in dynamic case. The acceptors (donors) are neutral or singly ionized. The neutral acceptors (donors) are *immobile*, the charged ones could be *mobile or almost immobile* [8].

### 3.1. Dynamic electromechanical response in decoupling approximation: flexoelectric, Vegard and electron-phonon contributions

Decoupling approximation has been recently used for MIECs [56, 67] and much earlier for the local electromechanical response calculations of ferroelectrics [88, 89, 90, 91, 92, 93, 94]. As the sort of perturbation approach, it consists of two successive stages. On the first stage one neglect the elastic stresses originated from electrostriction, piezoelectric effect (for ferroelectrics) and flexoelectric effects and Vegard expansion, in order to calculate of electric potential and mobile charges distribution. On the second stage all these effects are taken into account when the system strain is calculated using the electric potential and mobile charges distribution calculated on the first stage. The accuracy of decoupling approximation is surprisingly high even for ferroelectrics (error is proportional to the squire of the electromechanical coupling coefficient) and approved by other numerical methods like phase-field and FEM [95]. Earlier we studied the accuracy of decoupling approximation for paraelectric SrTiO$_3$ and proved that it is valid with several % accuracy at film thicknesses more than several screening radius and moderate applied voltages. Moreover, the decoupling approximation is valid with very high accuracy at arbitrary thicknesses and voltages after the substitution of LGD-expansion coefficient $\alpha(T)$ with $\alpha_P = \alpha(T) + 3\tilde{\alpha}_{11}\langle P \rangle^2$ in the electrostriction contribution term (see designations and Fig.2 in Ref.[67]).

We suppose that the total stress tensor $\sigma_{ij}(\mathbf{r})$ is linearly proportional to the Vegard contribution, electron-phonon contribution, flexoelectric and electrostriction contributions:

$$\sigma_{ij}(\mathbf{r}) = c_{ijkl}u_{kl}(\mathbf{r}) + \begin{pmatrix} \Xi_{ij}^{C}(n(\mathbf{r}) - n_0) + \Xi_{ij}^{V}(p(\mathbf{r}) - p_0) + \\ -\beta_{ij}^{a}(N_a^{-}(\mathbf{r}) - N_{a0}^{-}) - \beta_{ij}^{d}(N_d^{+}(\mathbf{r}) - N_{d0}^{+}) \end{pmatrix} - \gamma_{ijkl}\frac{\partial^2 \varphi}{\partial x_k \partial x_l} + \tilde{q}_{ijkl}\frac{\partial \varphi}{\partial x_k}\frac{\partial \varphi}{\partial x_l}. \quad (1)$$

Here $c_{ijkl}$ is the tensor of elastic stiffness, $u_{kl}(\mathbf{r})$ is the strain tensor, $\Xi_{ij}^{C,V}$ is a tensor deformation potential of electrons in the conduction (*C*) and valence bands (*V*), $\beta_{ij}^{a,d}$ are the *Vegard expansion* tensors for acceptors (donors). $N_d^{+}(\mathbf{r})$ is the instant concentration of mobile ionized donors, $N_a^{-}(\mathbf{r})$ is the instant concentration of mobile ionized acceptors, $N_{d0}^{+}$ and $N_{a0}^{-}$ are their stoichiometric equilibrium concentrations; $n(\mathbf{r})$ is the concentration of electrons in the conduction band, $p(\mathbf{r})$ is the concentration of holes in the valence band, $n_0$ and $p_0$ are their equilibrium concentrations, $\varphi(\mathbf{r})$ is the electric potential. Flexoelectric strain tensor $\gamma_{ijkl}$ has been measured experimentally for several substances and it was found to vary by several orders of magnitude from $10^{-11}$C/m to $10^{-6}$C/m [96, 97, 98].

$\tilde{q}_{ijkl}$ is the electrostriction tensor that couples stress and electric field. It is related with the electrostriction tensor $q_{ijkl}$, that couples strain and polarization, via the dielectric susceptibility $\chi_{ij} = \varepsilon_0(\varepsilon_{ij} - \delta_{ij})$ as $\tilde{q}_{ijkl} = \chi_{ip}\chi_{jq}q_{pqkl}$ ($\varepsilon_{ij}$ is the dielectric permittivity). In fact the electrostriction coefficients for typical *semiconductors with* low dielectric permittivity (*smaller than several tens*) are such that the electrostriction contribution becomes essential only at high electric fields (see e.g. [99, 100]). However for paraelectrics with high dielectric permittivity $\varepsilon_{ij}$, e.g. for SrTiO$_3$, electrostriction contribution can be dominant even at moderate electric fields [67].

Note, that Eq. (1) requires the reference crystallographic lattice to be defined, as analyzed for pure diffusion-stress coupling by Larche and Cahn [101]. The reference lattice is regarded strain-free for the case of zero electric potential: $\varphi = 0$ and therefore $n(\mathbf{r}) = n_0$, $p(\mathbf{r}) = p_0$, $N_a^{-}(\mathbf{r}) = N_{a0}^{-}$, $N_d^{+}(\mathbf{r}) = N_{d0}^{+}$.

Lame-type equation for the electromechanical displacement $u_i$ can be obtained from the equation of mechanical equilibrium $\partial \sigma_{ij}(\mathbf{r})/\partial x_j = 0$, where the stress tensor $\sigma_{ij}(\mathbf{r})$ is given by Eq.(1). Mechanical boundary conditions [102] corresponding to the ESM experiments [23] are defined on the mechanically free interface, $z = 0$, where the normal stress $\sigma_{3i}$ is absent, and on clamped interface $z = h$, where the displacement $u_i$ is fixed:

$$\sigma_{3i}(x_1, x_2, z = 0) = 0, \qquad u_i(x_1, x_2, z = h) = 0. \qquad (2)$$

Using the decoupling approximation in the 1D-Poisson equation for electric potential, mechanical displacement of the MIEC surface caused by the flexoelectric, electronic and ionic contributions was calculated as [56]:

$$u_3(z=0) \approx \int_0^h \frac{dz'}{c_{33}} \left( \begin{array}{c} \xi^C(n(z') - n_0) + \xi^V(p(z') - p_0) - \mu^a(N_a^-(z') - N_{a0}^-) \\ + \mu^d(N_d^+(z') - N_{d0}^+) + \tilde{q}_{33}\left(\frac{\partial\varphi(z')}{\partial z'}\right)^2 \end{array} \right) \qquad (3)$$

Here $e$ is the electron charge absolute value, constants $\xi^C \equiv \left(\Xi_{33}^C - \frac{\gamma_{3333} e}{\varepsilon_{33}^S \varepsilon_0}\right)$, $\xi^V \equiv \left(\Xi_{33}^V + \frac{\gamma_{3333} e}{\varepsilon_{33}^S \varepsilon_0}\right)$, $\mu^a \equiv \left(\beta_{33}^a + \frac{\gamma_{3333} e}{\varepsilon_{33}^S \varepsilon_0}\right)$ and $\mu^d \equiv \left(-\beta_{33}^d + \frac{\gamma_{3333} e}{\varepsilon_{33}^S \varepsilon_0}\right)$, $\varepsilon_{33}^S$ is MIEC dielectric permittivity, $\varepsilon_0$ is the universal dielectric constant.

From Eq.(3) that the MIEC surface displacement is proportional *to the total charge of each species.* Note, that the relation between the total charge and electrostatic potential on the semiconductor surface are well established [103]. Note, that the first terms in material constants $\xi^{C,V}$ and $\mu^{a,d}$ originated from the deformation potential or Vegard tensors, while the last ones originated from the flexoelectric coupling. Remarkably, that the strength of tensorial deformation potential $\Xi_{33}^{C,V}$ appeared comparable with Vegard tensor $\beta_{33}^{a,d}$ for correlated oxides (see **Table A1** in Supplementary materials,[104] **Appendix A**). Flexoelectric contribution is estimated in the **Table 1** of Ref. [56], and its value appeared comparable with Vegard contribution $\beta_{33}^{a,d}$ or even higher.

Note, that for ion-blocking electrodes the total number of ions remains the same, i.e. $\int_0^h (N_a^-(z) - N_{a0}^-) dz = 0$ and $\int_0^h (N_d^+(z) - N_{d0}^+) dz = 0$ when neglecting generation recombination effects. No such constrains exist for the case when one or two electrodes are ion conducting.

Note, that in principle the impedance spectroscopy formalism [78, 105, 106] can be used to derive the *linear* electromechanical response. However, for MIECs this approach requires the distributed models (e.g. see the Ref. [107] for detailed review), since lumped element models generally fail to reproduce the coupled electronic-ionic transport. This approach is then mathematically equivalent to the direct solution of coupled transport equations. Typically distributed circuit models also impose the electroneutrality condition, which is not the case at least near interfaces of MIEC thin films. However, in order to calculate the *nonlinear* electromechanical response in MIECs the impedance spectroscopy formalism should be modified to account for nonlinearity [108], since the impedance relations $V(t) \sim I(t) \cdot R(t)$ for I-V curves and mechanical displacement, as proportional to the total charge $Q(t)$ of each species, $u_3(t) \sim Q(t) \sim \int^t I(y)dy \sim \int^t V(y) \cdot R^{-1}(y)dy$, are valid in the time domain only. The convolution theorem for $u_3(\omega)$ should be applied in the spectral frequency domain, which breaks the proportionality $u_3(\omega) \sim 1/i\omega R$ allowing for the complex nonlinear temporal dependence of the impedance $R(t)$.

### *3.2. Poisson equations and electrodynamics boundary conditions*

For frequencies less then 1MHz, which is a typical operating limit for these experiments, the quasi-static approximation for electric field $E_z = -\dfrac{\partial \varphi(\mathbf{r})}{\partial z}$ works with high accuracy. Neglecting the flexoelectric term (decoupling approximation), the 1D-Laplace equation in the dielectric gap (if any) and the 1D-Poisson equation in MIEC film have the form:

$$\frac{d^2\varphi}{dz^2} = 0, \quad -H < z < 0 \quad \text{(gap)} \tag{4a}$$

$$\frac{d^2\varphi}{dz^2} = -\frac{\rho(\varphi)}{\varepsilon_{33}^S \varepsilon_0}, \quad 0 < z < h \quad \text{(MIEC film)} \tag{4b}$$

Here $\varphi(z)$ is the electric potential, $\varepsilon_{33}^S$ is MIEC dielectric permittivity. The charge density in MIEC film has the form:

$$\rho(z) = e(p(z) - n(z) - N_a^-(z) + N_d^+(z)) \tag{5}$$

The boundary conditions for the electrostatic potential $\varphi(z)$ are $\varphi(z = h) = 0$ on grounded substrate electrode), $\varphi(z = -H) = A_m/e + V$, on tip electrode-dielectric gap-film, $\varphi(z = +0) - \varphi(z = -0) \approx V_b$ on the tip electrode-dielectric gap-film, and

$$D_{2n} - D_{1n} = -\varepsilon_0 \left( \varepsilon_{33}^S \frac{\partial \varphi(z = +0)}{\partial z} - \varepsilon_{33}^g \frac{\partial \varphi(z = -0)}{\partial z} \right) = 0 \qquad (6)$$

on the film-dielectric gap boundary. Here, $A_m$ is the work function from the conducting tip electrode that typically determines the contact built-in potential $V_b$, $V(t) = V_0 \sin(\omega t)$ is the periodic voltage difference applied to the tip electrode at $z = -H$, $\varepsilon_{33}^g$ is the dielectric constant of the dielectric layer. The normal vector **n** is pointed from media 1 to media 2. The free surface charge is regarded absent at $z=0$ in dynamic case. Note, that the potential can be always set zero at the contact $z = h$, while the contact itself may either has contact barrier or be barrierless (ohmic).

### 3.3. Kinetic equations and boundary conditions

The total electric current is the MIEC film is $J_f = J_{Ds} + J_c$, where $J_{Ds}(z,t) = \varepsilon_0 \varepsilon_{33}^{S,g} \frac{\partial E_z}{\partial t}$ is the displacement current (existing both in the dielectric and in the MIEC), and $J_c(z,t)$ is the conductivity current that exists in the MIEC only. The continuity equation $\frac{\partial \rho}{\partial t} + \frac{\partial J_c}{\partial z} = 0$ should be solved along with the all electrodynamics equations.

The conductivity current $J_c(z,t) = \sum_{m=a,p,d,n} J_c^m$ consists of the acceptor ($J_c^a$), hole ($J_c^p$), donor ($J_c^d$) and electron ($J_c^n$) currents. Under negligibly small impact of the electron-hole recombination-generation process charges conservation equations are:

$$-\frac{\partial N_a^-}{\partial t} + \frac{1}{e}\frac{\partial J_c^a}{\partial z} = 0, \qquad (7a)$$

$$\frac{\partial p}{\partial t} + \frac{1}{e}\frac{\partial J_c^p}{\partial z} = 0, \qquad (7b)$$

$$\frac{\partial N_d^+}{\partial t} + \frac{1}{e}\frac{\partial J_c^d}{\partial z} = 0, \tag{7c}$$

$$-\frac{\partial n}{\partial t} + \frac{1}{e}\frac{\partial J_c^n}{\partial z} = 0. \tag{7d}$$

The electron and hole conductivity currents are proportional to the gradients of the carrier electrochemical potentials levels as $J_c^p = -e\eta_p p \frac{\partial \zeta_p}{\partial z}$, and $J_c^n = e\eta_n n \frac{\partial \zeta_n}{\partial z}$, where $\eta_{n,p}$ is the constant mobility of electrons (holes) and

$$-\zeta_p(z) \approx \Xi_{ij}^V u_{ij}(z) + e\varphi(z) + k_B T \ln\left(\frac{p(z)}{p_0}\right). \tag{8a}$$

Note that holes quasi-Fermi levels Eq. (8a) are typically defined in the BPN approximation, since they are mostly not degenerated in the MIECs.

At the same time, the electrons in the correlated oxides the well-localized. Using the rectangular DOS with constant density of states $g_n = N_n/(\delta E_n)$ over the range $[E_n, E_n + \delta E_n]$, electrochemical potential was derived as [109]:

$$\zeta_n \approx E_C + \Xi_{ij}^C u_{ij}(z) - e\varphi(z) - k_B T \ln\left(\frac{\exp(-\varepsilon n(z)/N_n) - \exp(-\varepsilon)}{1 - \exp(-\varepsilon n(z)/N_n)}\right). \tag{8b}$$

Here dimensionless energy $\varepsilon = \delta E_n/k_B T$, $k_B$=1.3807×10$^{-23}$ J/K, $T$ is the absolute temperature. In the case $\varepsilon n(z)/N_n \ll 1$ Eq.(8b) gives BPN approximation $\zeta_n \approx E_C + \Xi_{ij}^C u_{ij} - e\varphi + k_B T \ln(n/n_0)$, where the equilibrium concentration $n_0 = \frac{N_n}{\varepsilon}(1 - \exp(-\varepsilon))$.

Then, we substitute the acceptor and donor conductivity currents in Eq.(7) as proportional to the gradients of the corresponding electrochemical potentials [103]:

$$J_c^a = e\eta_a N_a^- \frac{\partial \zeta_a}{\partial z}, \quad \zeta_a(z) = -E_a + \beta_{jk}^a u_{jk}(z) - e\varphi(z) - k_B T \ln\left(\frac{N_a - N_a^-(z)}{N_a^-(z)}\right), \tag{8c}$$

$$J_c^d = -e\eta_d N_d^+ \frac{\partial \zeta_d}{\partial z}, \quad -\zeta_d(z) = E_d + \beta_{jk}^d u_{jk}(z) + e\varphi(z) - k_B T \ln\left(\frac{N_d - N_d^+(z)}{N_d^+(z)}\right) \tag{8d}$$

Here $\eta_{d,a}$ is the constant mobility of donors (acceptors), $E_{d,a}$ are their levels position calculated from the bottom of conductive band. Approximate equalities $\ln\left(\frac{N_a - N_a^-(z)}{N_a^-(z)}\right) \approx \ln\left(\frac{N_a}{N_a^-(z)}\right)$ and $\ln\left(\frac{N_d - N_d^+(z)}{N_d^+(z)}\right) \approx \ln\left(\frac{N_d}{N_d^+(z)}\right)$ correspond to the BPN approximation [110]. From Eq.(8c-d) one derives that $N_{a0}^- = N_a\left(1 + \exp\left(-\frac{E_F + E_a}{k_B T}\right)\right)^{-1}$ and $N_{d0}^+ = N_d\left(1 + \exp\left(\frac{E_F + E_d}{k_B T}\right)\right)^{-1}$ in the case $\varphi = 0$, when $\zeta_a = \zeta_d = E_F$.

Note that dynamics described by Eq. (8) should be valid for local thermal equilibrium conditions and in presence of small local gradients, i.e., $J_c^m \sim \eta_m \partial \zeta_m / \partial z$ even when the system is not in global equilibrium. For the high values of driving force, $\partial \zeta_m / \partial z$, one should use expression for currents derived by Riess and Maier [80].

Concentration dependences of ionized acceptors electrochemical potential are shown in **Fig. 2**. From the data in the plots the BPN approximation in Eq.(8c) works well for concentrations $N_a^- < (N_a^-)_{cr}$, where $(N_a^-)_{cr} = N_a$ as anticipated (see dashed lines). The condition $(N_a^-)_{cr} \approx N_a$ can be readily achieved in the vicinity of the film interfaces, where the space charge accumulation takes place. Exactly in the regions BPN approximation for acceptor electrochemical potential (and consequently the linear drift-diffusion model for their conductivity currents) become inapplicable.

The material boundary conditions relevant for the considered problem correspond to the limiting cases of the general Chang-Jaffe conditions [75, 76], namely

$$\left(J_c^p - w_{p0}(p - p_{S0})\right)\big|_{z=0} = 0, \quad \left(J_c^p + w_{ph}(p - p_{Sh})\right)\big|_{z=h} = 0, \quad (9a)$$

$$\left(J_c^n + w_{n0}(n - n_{S0})\right)\big|_{z=0} = 0, \quad \left(J_c^n - w_{nh}(n - n_{Sh})\right)\big|_{z=h} = 0, \quad (9b)$$

$$\left(J_c^a + w_{a0}(N_a^- - N_{a0}^-)\right)\big|_{z=0} = 0, \quad \left(J_c^a - w_{ah}(N_a^- - N_{ah}^-)\right)\big|_{z=0,h} = 0, \quad (9c)$$

$$\left(J_c^d - w_{d0}(N_d^+ - N_{d0}^+)\right)\big|_{z=0} = 0, \quad \left(J_c^d + w_{dh}(N_d^+ - N_{dh}^+)\right)\big|_{z=h} = 0. \quad (9d)$$

Upper and lower signs correspond to boundaries $z=h$ and $z=0$ respectively. Here $w_{p0,h}$, $w_{n0,h}$, $w_{a0,h}$ and $w_{d0,h}$ are positive rate constants of surface discharge [75, 111, 7] corresponding to boundaries $z=h$ and $z=0$ respectively; frequency-independent constants $p_{S0,h}$, $n_{S0,h}$, $N^{-}_{a0,h}$, $N^{+}_{d0,h}$ are equilibrium surface concentrations of holes, electrons, acceptors and donors at the film interfaces $z=h$ and $z=0$ respectively (at the absence on any currents). The conditions (9) contain the continuous transition from the **open conducting contacts** ($w_n \to \infty \Rightarrow n(\rho,0,t) = n_S$ and/or $w_d \to \infty \Rightarrow N^{\pm}_{d,a}(\rho,h,t) = N^{\pm}_{d0,h}$) to the **interface limited kinetics** ($w_{n,d} > 0$) and **blocking contacts** ($w_{n,d} = 0$) [111].

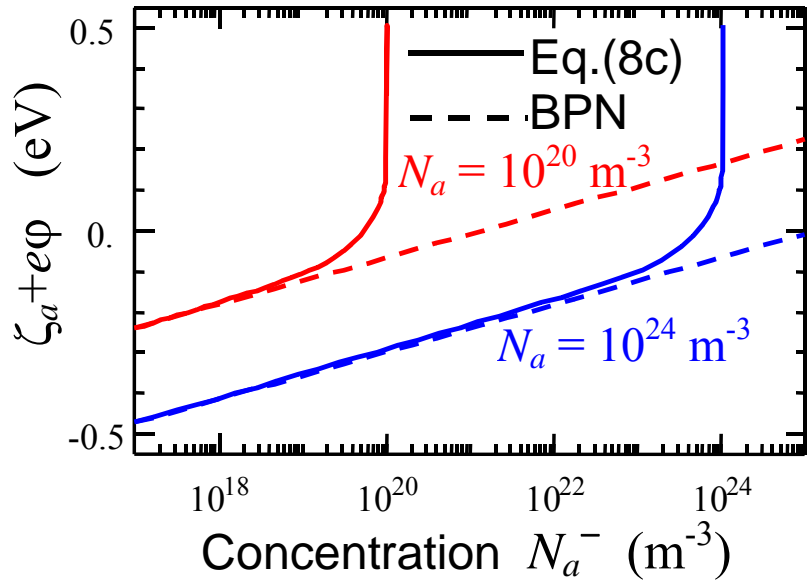

**Fig. 2.** Concentration dependences of ionized acceptors electrochemical potential levels calculated for $N_a = 10^{20}$ m$^{-3}$ and $10^{24}$ m$^{-3}$; $T = 293$ K, $u_{jk} = 0$. Solid and dashed curves represent the levels calculated from expressions (8c) and Boltzmann-Planck-Nernst approximation (BPN) correspondingly.

Equations (3)-(8) form the closed form nonlinear mixed boundary value problem, while the boundary conditions (9) are linear. The solutions for the case of periodic external voltage change $V(t) = V_0 \sin(\omega t)$) will be analyzed below in linear approximation analytically and for nonlinear case numerically.

## 4. Dynamic electromechanical response

### 4.1. Linear electromechanical response of MIEC film

Here we analyze the linear strain response caused by the periodic tip voltage for the *one prevailing type of carriers*, e.g. when acceptor mobility is absent or much smaller than the holes one, donors are almost absent and thus the concentration of the free electrons is also negligible in comparison with the holes concentration. Analytical expressions for the linear strain response caused by the periodic tip voltage was derived in decoupling approximation [Supplementary materials [104], **Appendix B]**. Mobile donors and electrons are not considered, but this can be done in similar way.

Substituting in Eq.(3) the expressions for charges variation $\delta Q_p(\omega) = \int_0^h (p(z) - p_0) dz$ (see Supplementary materials [104], **Appendix B**), we obtained an approximate analytical expression for the surface displacement of the MIEC film:

$$u_3(z=0,\omega) \approx -\xi^V \frac{\varepsilon_0 \varepsilon_{33}^S}{c_{33}} \cdot \frac{V_0}{h+\tilde{H}} \frac{(1-\exp(k(\omega)h))^2}{(1+\exp(2k(\omega)h))} \left( i\omega\tau_M + \frac{\tanh(hk(\omega)) + k(\omega)\tilde{H}}{(h+\tilde{H})k(\omega)} \right)^{-1} \quad (10a)$$

Here $\tilde{H} = \frac{\varepsilon_{33}^S}{\varepsilon_{33}^g} H$ and relaxation time $\tau_M = \frac{\varepsilon_0 \varepsilon_{33}^S}{e p_0 \eta_p}$, $\omega$ is the frequency of the voltage $V(t) = V_0 \exp(i\omega t)$ applied to the SPM probe. The spatial scale

$$k(\omega) = \sqrt{\frac{i\omega}{D} + \frac{1}{R_S^2}} \equiv \frac{1}{R_S}\sqrt{i\tau_M \omega + 1} \quad (10b)$$

is defined by the diffusion coefficient $D = \eta_p \frac{k_B T}{e}$ and Debye screening radius

$$R_S = \sqrt{\frac{\varepsilon_{33}^S \varepsilon_0 k_B T}{e^2 p_0}}.$$

In the approximate expression (10a) we should neglect electrostriction, because in the linear approximation and under the absence of built-in static electric field, electrostriction response will be manifested at frequencies doubled in comparison with the frequency of excitation electric field. In the linear regime double-frequency signal can be excluded experimentally. Analytical results obtained from Eqs.(10) are justified, if the hole conductivity is constant proportional to the mobility and average concentration of holes, while acceptor conductivity is negligibly small in comparison with the hole one. Otherwise numerical modeling of the nonlinear problem should be performed.

Note, that Eqs.(10) is derived for the case $J_\omega^c(0)=0$, $\rho_S(h)=0$, where the space charge density $\rho_S(h)=e(-N_a^-(h)+p(h))$. Other types of the boundary conditions, e.g. $\rho_S(0)=\rho_S(h)=0$ and $J_\omega^c(0)=J_\omega^c(h)=0$, lead to the total charge absence and consequently to zero surface displacement $u_3(\omega)=0$ in the linear decoupling approximation (see **Table C.1** in Supplementary materials [104]).

**Figures 3** represent the frequency spectra of the surface displacement (10) for several gap thicknesses and mixed-type boundary conditions $J_\omega^c(0)=0$, $\rho_S(h)=0$. The displacement $u_3(z=0,\omega)$ is proportional to the total space charge $\delta Q_p(\omega)=\int_0^h (p(z)-p_0)dz$, since the total acceptor charge $\delta N_a^-(\omega)=\int_0^h (N_a^-(z)-N_{a0}^-)dz \equiv 0$ is zero for the ion-blocking boundary conditions (9a). In the limiting case of zero gap ($H\to 0$) the displacement is maximal; it decreases with the gap thickness increase. The total displacement absolute value monotonically decreases with frequency increase; while its imaginary part has maximum.

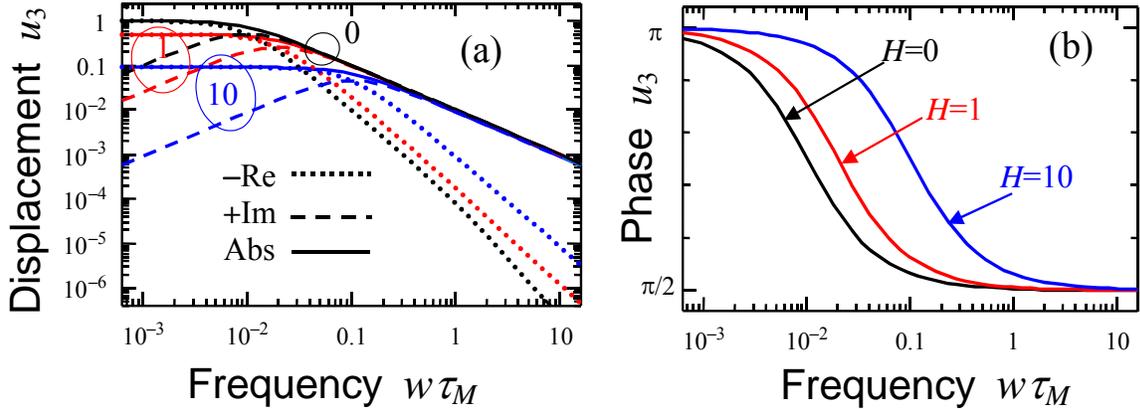

**Fig. 3.** (a) Real, imaginary parts, absolute value (dotted, dashed and solid curves) and phase (b) of the normalized surface displacement $u_3(0,\omega)$ vs. dimensionless frequency $w=\omega/2\pi$ calculated for several gap thickness $\tilde{H}/R_S = 0, 1, 10$ (figures near the curves). Film thickness $h/R_S = 100$, mixed boundary conditions $J_\omega^c(0)=\rho_S(h)=0$ are imposed.

*4.2. Nonlinear dynamic strain–voltage response of MIEC*

Below we analyze the dynamic electromechanical response caused by the mobile ionized donors and electrons in the ionic semiconductor film. Note, that the dynamic electromechanical response caused by the mobile ionized acceptors and holes can be analyzed in a similar way.

For numerical modeling we introduce the Debye screening radii

$$R_S = \sqrt{\frac{\varepsilon_{33}^S \varepsilon_0 k_B T}{e^2 n_0}}, \quad (11a)$$

Maxwellian relaxation time that determines the timescale of the considered problem is

$$\tau_M = \frac{R_S^2 e}{\eta_n k_B T}, \quad (11b)$$

Thus below we operate with dimensionless frequency $\tau_M f$, where linear frequency $f = \omega/2\pi$.

Dimensionless rate constants

$$\widetilde{w}_{n0,h} = \frac{\tau_M}{R_S e} w_{n0,h}, \quad \widetilde{w}_{d0,h} = \frac{\tau_M}{R_S e} w_{d0,h}. \quad (11c)$$

and dimensionless electromechanical response of electrons or donors:

$$u_3(z=0) \approx \frac{1}{2} \int_0^h \frac{d\tilde{z}}{c_{33}} \left( \xi^C \left( \frac{n(\tilde{z}) - n_0}{n_0} \right) + \mu^d \left( \frac{N_d^+(\tilde{z}) - N_{d0}^+}{n_0} \right) + \frac{\tilde{q}_{33} k_B T}{\varepsilon_{33}^S \varepsilon_0} \left( \frac{\partial \widetilde{\varphi}(\tilde{z})}{\partial \tilde{z}} \right)^2 \right) \quad (11d)$$

Here $\tilde{z} = z/R_S$. Electron and donor contributions in Eq.(11d) are divided by the factors $c_{33}/\xi^C$ and $c_{33}/\mu^d$ correspondingly, as compared with Eq.(3). The dimensionless parameters $\tilde{q}_{33} k_B T / (2\varepsilon_{33}^S \varepsilon_0 \xi^C)$ and $\tilde{q}_{33} k_B T / (2\varepsilon_{33}^S \varepsilon_0 \mu^d)$ determines the relative strength of electrostriction contribution. Other dimensionless variables used in Eq.(3)-(9) under the simulations are introduced in Supplementary materials [104], **Appendix C1.** Also there we analyzed some typical I-V curves in the **Appendix C2**.

Dimensionless electromechanical response (11d) was calculated numerically with the help of Matlab [112] for the external voltage frequency range $\tau_M f = 0.001 - 0.1$ and different types of boundary conditions (9). The *pdepe* function was used, the latter solved the nonlinear problem via second order finite difference in space and up to 5$^{th}$ order numerical differentiation formula in time. The mesh was nonuniform with up to 1000 elements and the solution time steps were chosen adaptively according to the ode15s algorithm. Dimensionless electrochemical potentials (Slootblom formulation [113]) were used as variables for the solution to ensure the stability of the numerical problem. Typical response curves $\tilde{u}_3(V, f)$ are shown in Figs. 4-9.

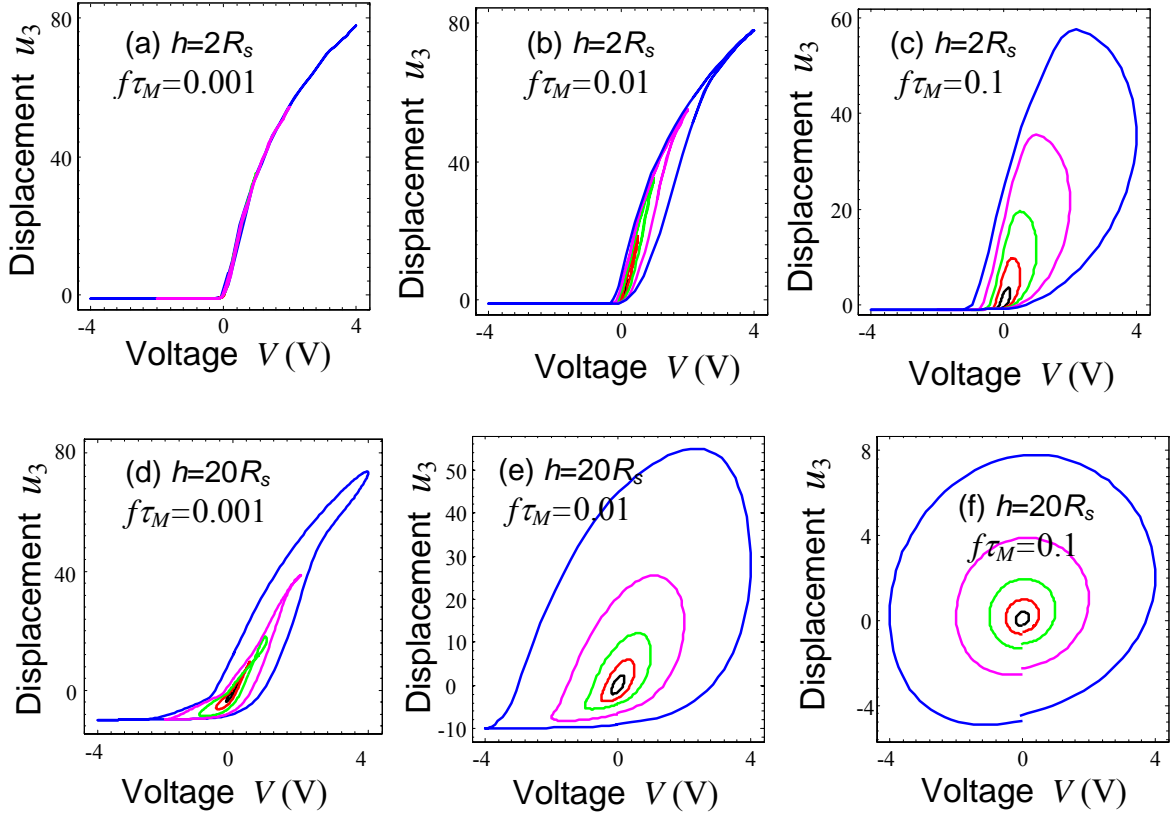

**Fig. 4.** Electromechanical response $\tilde{u}_3(V, f)$ calculated for different frequencies: $\tau_M f = 0.001$ (a, d), $\tau_M f = 0.01$ (b, e), and $\tau_M f = 0.1$ (c, f). Film thickness $h/R_s = 2$ (a, b, c) and $h/R_s = 20$ (d, e, f). Interface z=0 is almost electron blocking, $J_c^n(0) = 0$ (we put $\tilde{w}_{n0} \leq 10^{-2}$), interface z=h is almost electron conducting (we put $\tilde{w}_{nh} \geq 10^2$). Both interfaces are ion blocking: we put $w_{d0,h} = 0$ to reach $J_c^d(0) = J_c^d(h) = 0$. Band structure parameters: $E_n$=0 eV, $\delta E_n$=0.5 eV for electrons and $E_d$=0.1 eV for donors. Equilibrium surface concentrations are assumed to be equal to the bulk ones, full amounts ratio $N_d/N_n = 0.1$, mobilities ratio $\eta_d/\eta_n = 0.1$. Also we neglected electrostriction contribution, $\tilde{q}_{33}$=0.

Different loops (black, red, green and blue ones) in each plot correspond to the increasing voltage amplitude $V_0$ (in volts). All plots are generated using expressions (8) for the chemical potential of carriers. The differences in loop shape mainly originate from the type of boundary conditions, external voltage frequency and film thickness as discussed in the subsections 4.2.1-3.

In **Figures 4, 5, 7** and **8** we neglect the electrostriction impact into the electromechanical response (possible case of dielectrically linear materials, like yttria-stabilized zirconia, $LiCoO_2$, $LiMn_2O_4$, $LiC_6$). Electrostriction contribution is included in **Fig. 6, 9** and **10** for material parameters $\xi^C$=10 eV, $\mu^d$=10 eV (recalculated from known flexoelectric coefficients and the data of Ref. [114]), $q_{33}= -13.7 \cdot 10^9$ m J/C$^2$ and $\varepsilon_{33} = 300$ corresponding to $SrTiO_3$ with oxygen vacancies. Since the oxygen vacancy concentration (and corresponding conductivity) can be tuned in the wide range for $SrTiO_3$ [115, 116], we cannot define $\tau_M$ for all cases, but rather consider the range $\tau_M f = 0.001 - 0.1$.

### 4.2.1. Ion-blocking and electron-conducting interfaces

The hysteresis-like loops, shown in **Figs. 4**, are calculated for the case of asymmetric mixed-type electronic boundary conditions (9): interface $z=0$ is **almost** electron blocking ("almost" means that results remained the same when we put $\tilde{w}_{n0} \leq 10^{-2}$ in Eq.(9b)), interface $z=h$ is almost electron conducting (we put $\tilde{w}_{nh} \geq 10^2$); both interfaces are ion blocking: we put $w_{d0,h} = 0$ to reach $J_c^d(0) = J_c^d(h) = 0$. Different loops (black, red, green and blue ones) correspond to the different values of maximal voltage $V_0$. Plots (a, b, c) are generated for thin film ($h=2R_S$) and plots (d, e, f) for thicker ones ($h=20R_S$). The loop shape is quasi-ellipsoidal only at small voltage amplitudes $V_0 < k_B T/e$ and becomes asymmetric hysteresis-like with $V_0$ increase for $f \tau_M \leq 0.01$. The loops becomes noticeably open (or even circle-like) with the frequency increase $f \tau_M \geq 0.01$. The loop opening becomes much stronger with the thicknesses increase. Note, that the response curves are strongly asymmetric with respect to the voltage sign $V \rightarrow -V$, as can be expected from the asymmetry of the interface electronic conductivity. We further emphasize that the donor blocking boundary conditions ($J_c^d(0) = J_c^d(h) = 0$) and negligible generation-recombination effects, the continuity equation rules that $\dfrac{d}{dt}\int_0^h N_d^+(z)dz = 0$ and ionized donors contribute nothing to the response $u_3(V,f)$. Thus, only the total changes of the electron amount contribute into the MIEC film surface displacement.

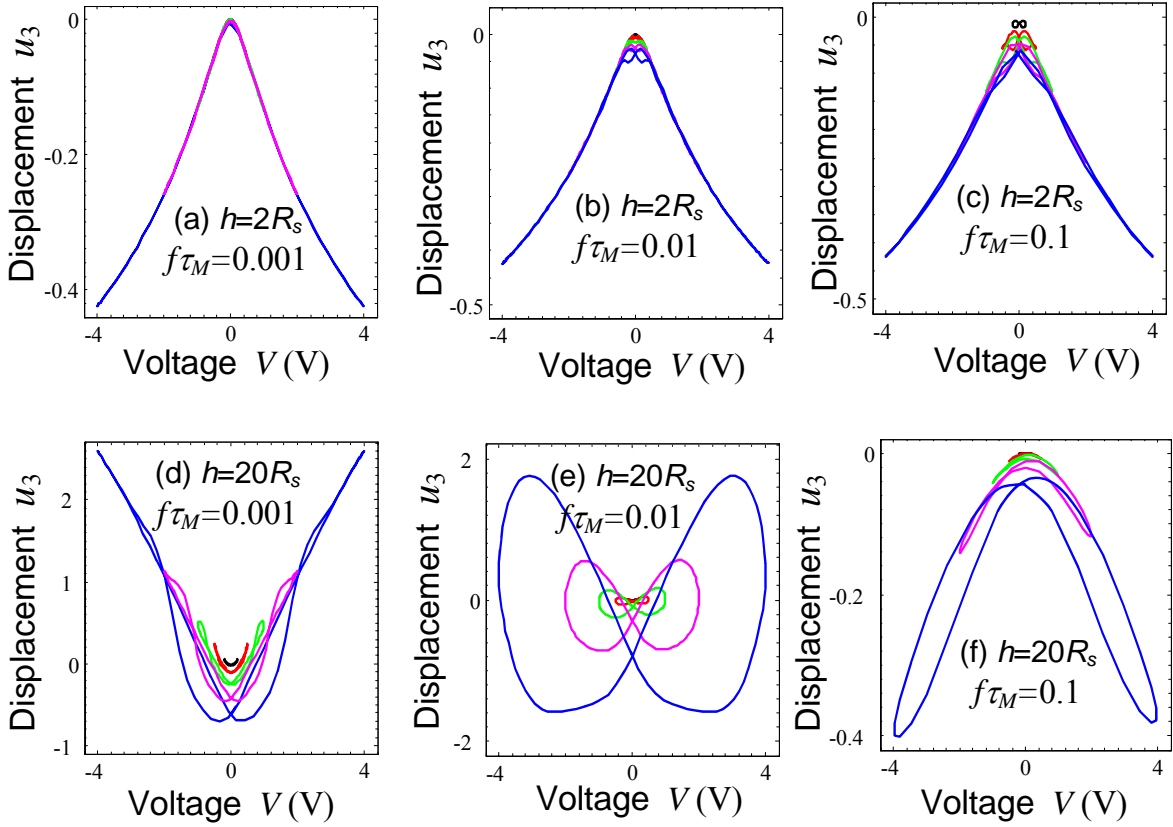

**Fig. 5.** Electromechanical response $\tilde{u}_3(V,f)$ calculated for different frequencies: $\tau_M f = 0.001$ (a, d), $\tau_M f = 0.01$ (b, e), and $\tau_M f = 0.1$ (c, f). Film thickness $h/R_s = 2$ (a, b, c) and $h/R_s = 20$ (d, e, f). Interfaces z=0 and z=h are almost electron conducting (we put $\tilde{w}_{n0,h} \geq 10^2$). Both interfaces are ion blocking: we put $w_{d0,h} = 0$ to reach $J_c^d(0) = J_c^d(h) = 0$. Other parameters are listed in the capture to **Fig. 4**.

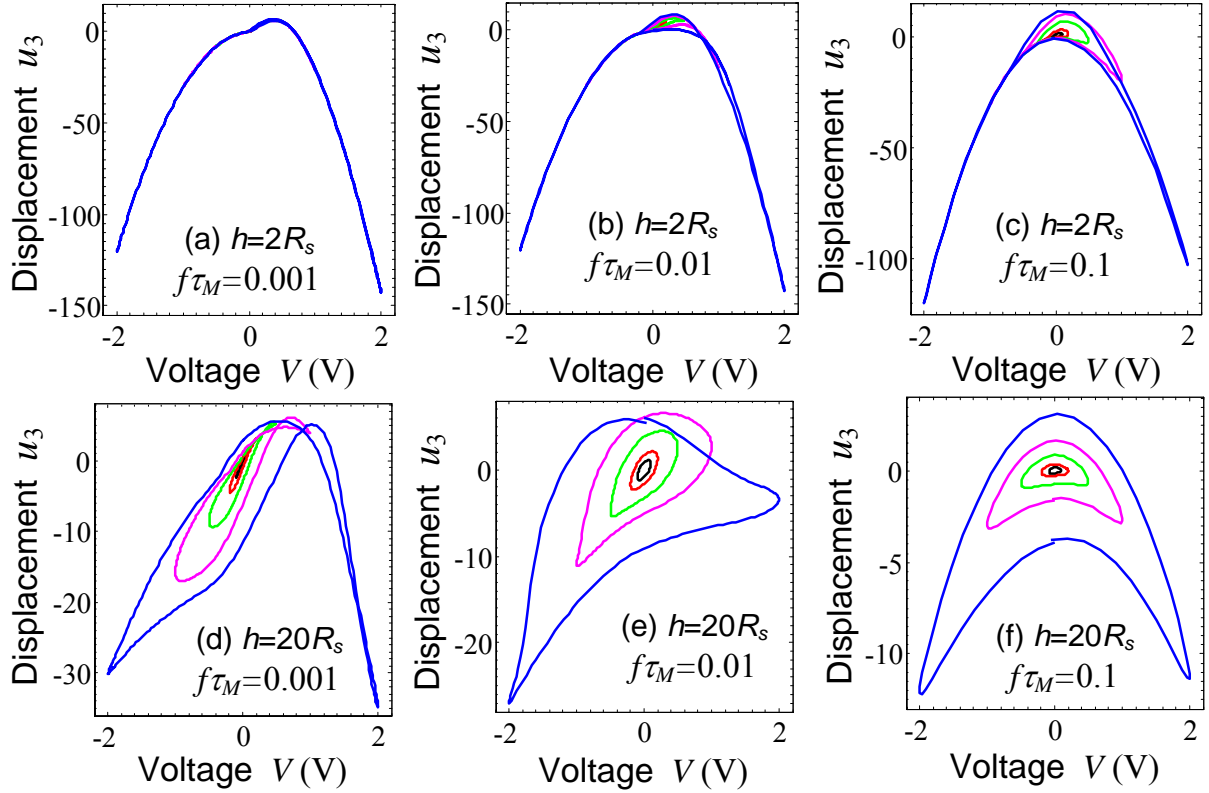

**Fig. 6.** Electromechanical response $\tilde{u}_3(V,f)$ calculated for nonzero electrostriction coefficient $\tilde{q}_{33} k_B T / (2\varepsilon^S_{33} \varepsilon_0 \xi^C) = -0.04$ at different frequencies: $\tau_M f = 0.001$ (a, d), $\tau_M f = 0.01$ (b, e), and $\tau_M f = 0.1$ (c, f). Film thickness $h/R_S = 2$ (a, b, c) and $h/R_S = 20$ (d, e, f). Boundary conditions and other parameters are listed in the capture to **Fig. 4**.

The response curves $u_3(V, f)$, shown in **Figs. 5**, are symmetric with respect to the voltage sign $V \to -V$, since the curves are calculated for the case of symmetric electron conducting and ion-blocking interfaces at z=0 and z=h. Note, that for the case the gaps should be absent. Different loops (black, red, green and blue ones) correspond to the different values of maximal voltage $V_0$. Plots (a, b, c) are generated for thin film ($h=2R_S$) and plots (d, e, f) for thicker ones ($h=20R_S$). The curves calculated for low frequencies $\tau_M f = 0.001$-0.01 are symmetric with respect to the voltage sign even after the first cycling. The curves generated at higher frequencies $\tau_M f = 0.1$ become symmetric with respect to the voltage sign only after relatively long relaxation of the initial conditions. The curves calculated for thick films are more overblown in comparison with the ones calculated for thin films (compare plots a, b, c with d, e, f). Finally, note that the nonlinear electromechanical response is absent for the completely blocking conditions $J_c^n(0) = J_c^n(h) = J_c^d(0) = J_c^d(h) = 0$.

Electrostriction is chosen negligibly small in **Figs. 4 and 5**, that corresponds to the case $\left|\tilde{q}_{33} k_B T / \left(2\varepsilon_{33}^S \varepsilon_0 \xi^C\right)\right| << 10^{-3}$. Electromechanical response $\tilde{u}_3(V, f)$ calculated for the same parameters as in **Fig. 4** and SrTiO$_3$ electrostriction coefficient $\tilde{q}_{33}$ is shown in **Figs. 6**. It is seen from the **Figs. 6** that electrostriction contribution to dynamical electromechanical response is of the same order or essentially higher than the Vegard contribution for paraelectrics and incipient ferroelectrics like SrTiO$_3$ due to high dielectric permittivity. Corresponding responses acquire "parabolic-like" and "moon-like" shape. Since the "parabolic-like" curves were calculated analytically for the static local electromechanical response of SrTiO$_3$ [67], the dynamical response calculated numerically tends to the static limit with the frequency decrease as anticipated. The hysteresis loop opens under the frequency increase (compare **Figs. 6a, 6b, 6c**). The film thickness increase leads to the electric field decrease and thus electrostriction contribution decreases (compare **Figs. 6a-c** with **6d-f**). Dependencies in **Figs. 6** are asymmetric with respect to the voltage sign due to the imposed symmetric mixed-type electronic boundary conditions.

### 4.2.2. Ion-conducting and electron-blocking interfaces

Here, we compare asymmetric and symmetric ion-conducting boundary conditions (9d). Both interfaces are electron blocking: we put $w_{n0} = w_{nh} = 0$ to reach $J_c^n(0) = J_c^n(h) = 0$. The hysteresis-like loops, shown in **Figs. 7**, are calculated for the case of asymmetric mixed-type ionic boundary conditions: interface z=0 is almost donor blocking, interface z=h is almost donor conducting; both interfaces are electron blocking. Different loops (black, red, green and blue ones) correspond to the different values of maximal voltage $V_0$. Plots (a, b, c) are generated for thin film ($h=2R_S$) and plots (d, e, f) for thicker ones ($h=20R_S$). At low frequencies $f \tau_M \leq 0.01$ the response curves are strongly asymmetric with respect to the voltage sign $V \to -V$ as anticipated from the asymmetry of the interfaces ionic conductivity. For the case only the total changes of the ionized donor amount contribute into the MIEC film surface displacement.

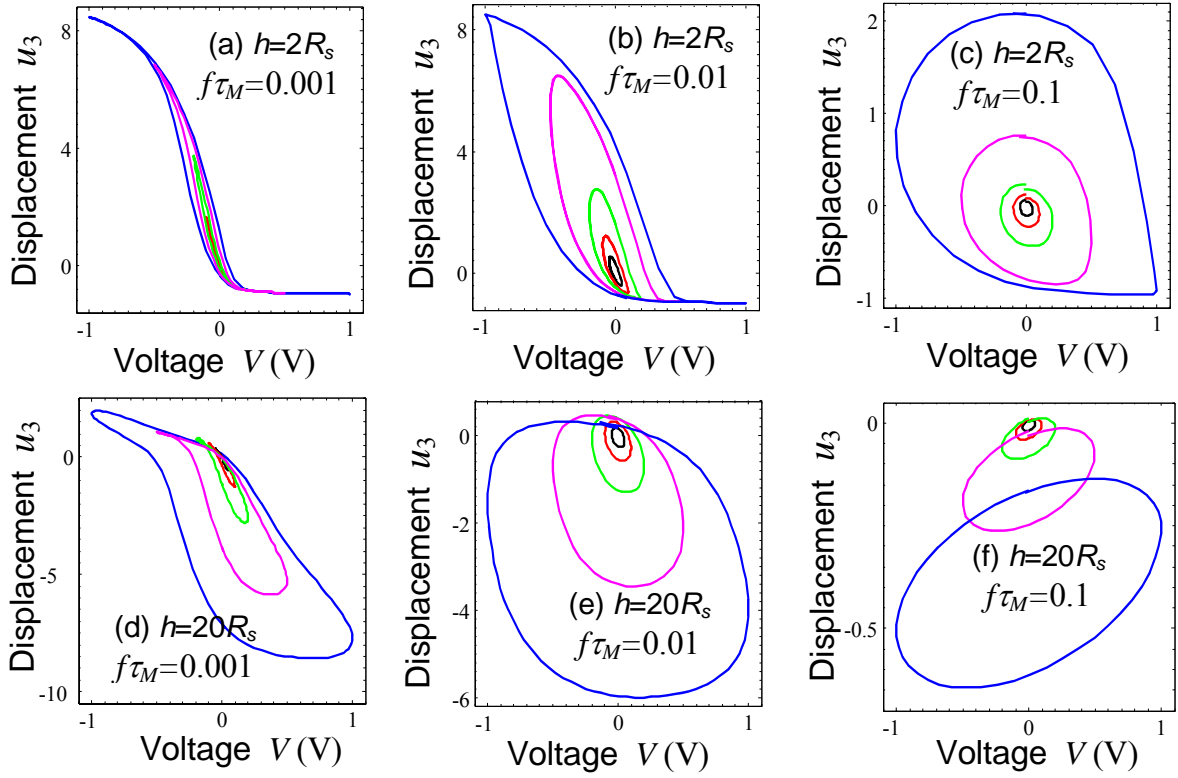

**Fig. 7.** Electromechanical response $\tilde{u}_3(V, f)$ calculated for different frequencies: $\tau_M f = 0.001$ (a, d), $\tau_M f = 0.01$ (b, e), and $\tau_M f = 0.1$ (c, f). Film thickness $h/R_S = 2$ (a, b, c) and $h/R_S = 20$ (d, e, f). Interface z=0 is almost donor blocking (we put $\tilde{w}_{d0} \leq 10^{-2}$ to reach $J_c^d(0) \approx 0$), interface z=h is almost donor conducting (we put $\tilde{w}_{dh} \geq 10^2$). Both interfaces are electron blocking: we put $w_{n0} = w_{nh} = 0$ to reach $J_c^n(0) = J_c^n(h) = 0$. Other parameters are listed in the capture to **Fig. 4**.

The loops become noticeably open and almost symmetric with the frequency increase $f \tau_M \geq 0.01$. The inflation becomes much stronger with the thicknesses increase. From **Figs. 7** the loop shape is elliptic for small voltages $V_0 < k_B T/e$, and the corresponding parameters depend on the film thickness and boundary conditions, which is consistent with analytical results of the subsection 4.1. For high maximal voltage $V_0$ the loop shapes demonstrate a pronounced size effect: the transition from the slim hysteresis to ellipse appears with the film thickness increase. The transition most probably originates from the acting electric field decrease with the film thickness increase: the thicker is the film the more close to linear is its response.

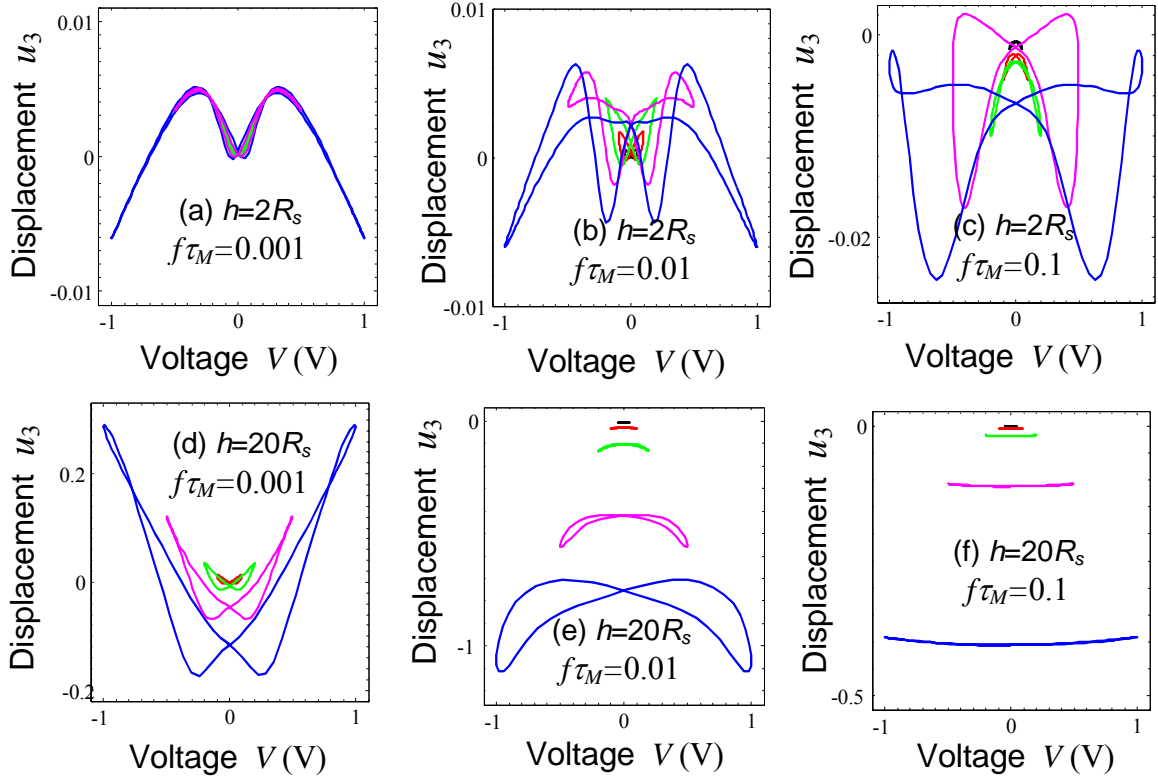

**Fig. 8.** Electromechanical response $\tilde{u}_3(V,f)$ calculated for different frequencies: $\tau_M f = 0.001$ (a, d), $\tau_M f = 0.01$ (b, e), and $\tau_M f = 0.1$ (c, f). Film thickness $h/R_S = 2$ (a, b, c) and $h/R_S = 20$ (d, e, f). Interfaces z=0 and z=h are almost donor conducting (we put $\tilde{w}_{dh} \geq 10^2$). Both interfaces are electron blocking: we put $w_{n0} = w_{nh} = 0$ to reach $J_c^n(0) = J_c^n(h) = 0$. Other parameters are listed in the capture to **Fig. 4**.

The response curves $u_3(V,f)$, shown in **Figs. 9**, are symmetric with respect to the voltage sign $V \to -V$, since the curves are calculated for the case of symmetric ion conducting and electron blocking interfaces at z=0 and z=h. Note, that for the case the gaps should be absent. Different loops (black, red, green and blue ones) correspond to the different values of maximal voltage $V_0$. Plots (a, b, c) are generated for thin film (h=5$R_S$) and plots (d, e, f) for thicker ones (h=20$R_S$). The curves calculated for low frequencies $\tau_M f = 0.001$-0.01 are symmetric with respect to the voltage sign even after the first cycling. The butterfly-like curves generated at higher frequencies $\tau_M f = 0.1$ become symmetric with respect to the voltage sign only after relatively long relaxation of the initial conditions (compare plots a, b, c with d, e, f).

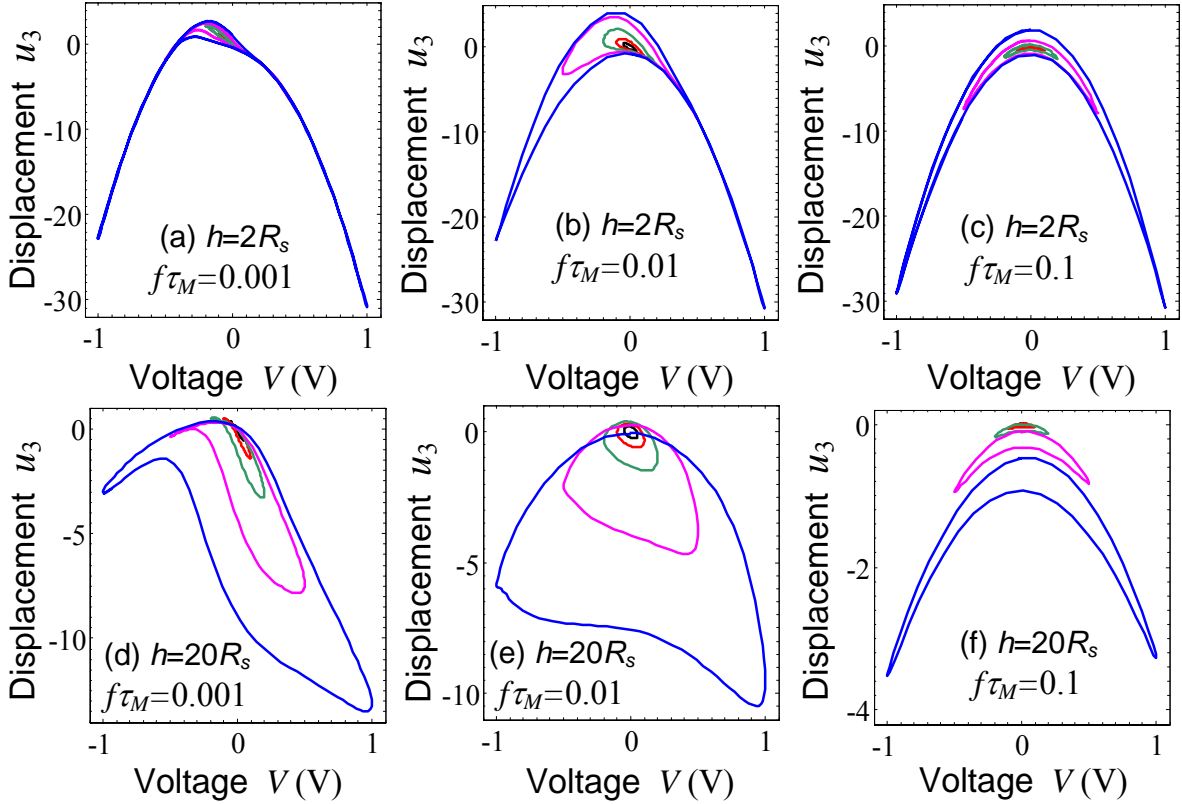

**Fig. 9.** Electromechanical response $\tilde{u}_3(V, f)$ calculated for nonzero electrostriction coefficient $\tilde{q}_{33} k_B T / (2\varepsilon_{33}^S \varepsilon_0 \mu^d) = -0.04$ at different frequencies: $\tau_M f = 0.001$ (a, d), $\tau_M f = 0.01$ (b, e), and $\tau_M f = 0.1$ (c, f). Film thickness $h/R_S = 2$ (a, b, c) and $h/R_S = 20$ (d, e, f). Boundary conditions and other parameters are listed in the capture to **Fig. 7**.

Electrostriction contribution is chosen negligibly small in **Figs. 7** and **8**, namely we regard $\left| \tilde{q}_{33} k_B T / (2\varepsilon_{33}^S \varepsilon_0 \mu^d) \right| << 10^{-3}$ when calculate the plots. Dynamical response $\tilde{u}_3(V, f)$ calculated for SrTiO$_3$ electrostriction coefficient $\tilde{q}_{33}$, asymmetric and symmetric ion-conducting boundary conditions are shown in **Figs. 9** and **10** correspondingly. **Figures 9** and

**10** demonstrate that electrostriction contribution is of the same order or even 1-2 orders higher than the ionic and electronic contributions. Corresponding responses acquire "parabolic-like" shape at low frequencies in thin films. The moon-like or asymmetric hysteresis loop opens under the frequency increase. The film thickness increase leads to the electric field decrease and thus electrostriction contribution decreases (compare with **Fig. 6**).

Quantitatively, the difference in the boundary conditions leads to asymmetry of the discrepancies and asymmetry of the loops shape, which correlates with results of the subsections 4.2.1. The main difference between the case of ion-blocking boundary conditions considered in subsections 4.2.1 and the ion-conducting top electrode considered in the subsection is the inverse loop orientation as anticipated from the substitution of the carrier charge electrons → donors. Similar effect can be expected for holes → acceptors.

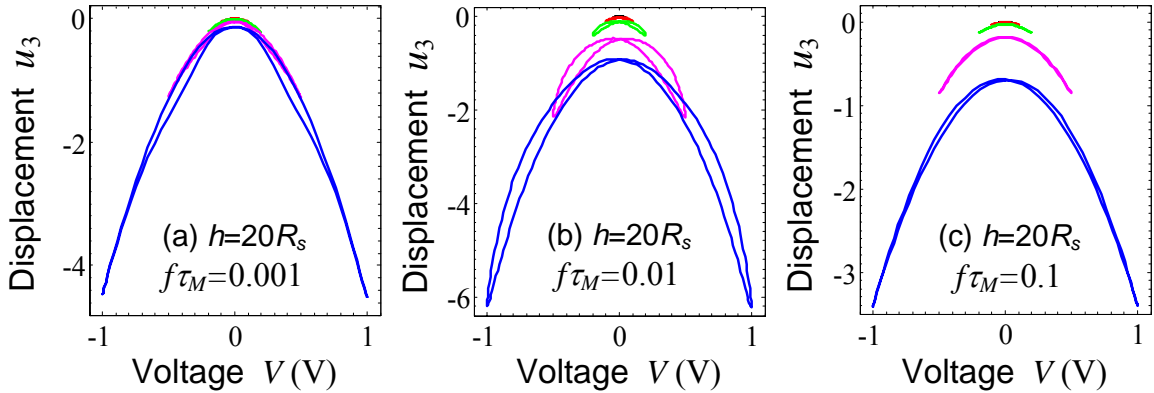

**Fig. 10.** Electromechanical response $\tilde{u}_3(V, f)$ calculated for nonzero electrostriction coefficient $\tilde{q}_{33} k_B T / (2\varepsilon_{33}^S \varepsilon_0 \mu^d) = -0.04$ at different frequencies: $\tau_M f = 0.001$ (a), $\tau_M f = 0.01$ (b), and $\tau_M f = 0.1$ (c). Film thickness $h/R_S = 20$. Boundary conditions and other parameters are listed in the capture to **Fig. 8**.

We expect that observable dynamical electromechanical response of MIECs should strongly depend on the relative strength of ionic, electronic and electrostriction contributions and boundary conditions type (carriers-blocking, carriers-conducting or mixed). In principle all regimes considered in the paper can be realized for proper electrodes (carriers-blocking, carriers-conducting or mixed). However, it is worth to underline that parabolic-like or moon-like shape is typical for the majority of loops in **Figs.6, 9** and **10** calculated for SrTiO$_3$. So, we may conclude that that dynamic electromechanical response of paraelectrics and incipient ferroelectrics like SrTiO$_3$ with oxygen vacancies or other mobile charge defects is primary determined by the strong electrostriction contribution and secondary by the electrode type.

**5. Summary remarks**

We performed analytical and numerical calculations of the ***dynamic*** electromechanical response of the MIEC film caused by the local changes of ions (acceptors or donors) concentration (*conventional stoichiometry contribution*); free electrons (holes) concentration (*electron-phonon coupling via the deformation potential*) and *flexoelectric effect*. Dynamic electromechanical response was not calculated previously, while our estimations performed for correlated oxides show that strength of all three contributions appeared comparable. Moreover, the coupling contribution proportional to the deformation potential may be stimulated by the local Jahn-Teller distortion existing in correlated oxides like La$_{1-x}$Sr$_x$MnO$_3$ and La$_{1-x}$Sr$_x$CoO$_3$. This allows relating the calculated electromechanical response with the local deformation potential of correlated oxides.

A great variety of possible nonlinear dynamic electromechanical response of MIEC films is predicted. Electromechanical responses mimic hysteresis loops with pronounced memory window and butterfly-like loops for partially and completely on-conducting boundary conditions correspondingly. Predicted strain-voltage hysteresis of piezoelectric-like, parabolic-like, moon-like and butterfly-like shape requires experimental justification in ionic semiconductors like correlated oxides, strontium titanate and resistive switching materials. Consequently, the SPM measurements of the MIEC film surface displacement could provide important information about the local oxidation level, electron-phonon interactions via the deformation potential and even Jahn-Teller distortions in the films.

**Acknowledgements**
A.N.M. and E.A.E. gratefully acknowledge multiple discussions with Prof. N.V. Morozovskii, Prof. A.K. Tagantsev and useful remarks given by Dr. Liangjun Li.

A.N.M., E.A.E. and G.S.S. acknowledge Grants of State Fund of Fundamental Research № UU30/004 and President Grant № GP/F32/099. L-Q.C. research is sponsored by the National Science Foundation (Materials World Network, DMR-0908718). FC thanks HKUST for providing the start-up funds. E.A.E. and A.N.M. further acknowledge user agreement with CNMS № UR-08-869. S.V.K. was supported by the U.S. Department of Energy, Basic Energy Sciences, Materials Sciences and Engineering Division.